# Plasmonic ultra-broadband polarizers based on Ag nano wire-slit arrays

Chunrui Han and Wing Yim Tam*

*Department of Physics and William Mong Institute of Nano Science and Technology*

*The Hong Kong University of Science and Technology*

*Clear Water Bay, Kowloon, Hong Kong, China*

## Abstract

We propose ultra-broadband reflective and absorptive polarizers in the visible range using multi-scaled Ag nano wire-slit arrays. The nano arrays can be tuned from reflective to absorptive by incorporating Ag wires/strips with different lengths/widths. The ultra-broadband nature of the absorptive array, with averaged absorption as high as ~80%, is due to the collective excitations of plasmonic resonances in the Ag wires/strips with different length scales. The Ag nano wire-slit arrays are realized experimentally by using a simple two-times shadowing vapor deposition method. They exhibit broadband transmission difference, in good agreement with simulations. The multi-scaled Ag nano wire-slit array has potential applications as broadband linear polarizers and anti-reflective materials in both optics and photovoltaics.

* Corresponding Author: phtam@ust.hk

Plasmonic metamaterials are artificial metallic nanostructures, with feature sizes smaller than the wavelength of interest, engineered to achieve unusual optical responses that are not available in nature [1, 2]. These properties are derived from localized surface plasmon resonances (LSPRs), i.e. collective oscillations of the conduction electrons on the surfaces of the nanostructures in resonance with the incident light [3]. LSPR nanostructures exhibit strong absorptions and enhancements of local electromagnetic fields leading to many fascinating applications, such as absorbers [4], light concentrators [5], optical sensors [6], antennae [7], and even in near-field scanning optical microscopy [8]. LSPR depends strongly on the sizes, shapes, orientations, and surrounding dielectric environment of the metallic nanostructures. Thus, different nanostructures can been designed to optimize the LSPR signal at one or multiple frequencies such as in metallic disks [9, 10], squares [11], T-shaped arrays [12, 13], nano-strips [14], nanoparticles [15] and etc., with either single size/layer or multi-sizes/layers. While it is relatively easy to control the resonance and bandwidth of single LSPR signal, it is still challenging to design broadband responses using multi-scaled/layered nanostructures, especially for the visible range. Moreover, for 2D structures with multi-scaled metallic elements on a surface, the magnitude of individual resonance is still small due to the limit density of elements [10]. Even though the magnitude of the resonances can be improved by stacking more layers of the 2D structures, the thick metallic components will degrade the overall transmission and thus limits them for mostly absorptive applications [14].

Recently, the resonant property of the metallic nanostructures has been extended to design optical devices such as wave plates [16-19] and polarizers [20-23]. For polarizers, they can be categorized into absorptive and reflective type. A typical absorptive polarizer, like the polaroid material made of polyvinyl alcohol plastic with an iodine doping [24], can absorb/dissipate light for polarization parallel to the polymer chains and thus transmit only the orthogonally polarized light. In contrast, reflective polarizer, e.g. parallel conducting wires [25], relies on the high reflectance of incident light for polarization parallel to the wires and thus also transmit also only the orthogonally polarization light. Moreover, a recent study reveals that the broadband transmission of linearly polarized light with electric field perpendicular to the metallic wires is due to the significantly reduced impedance mismatch by reducing the slit (Fabry-Perot) resonance in the metallic nano-slit array [26]. As a result, by controlling the anisotropy of the nanostructure, it is easy to achieve high and flat transmission for polarization perpendicular to the slit direction as well as reflection for the orthogonal direction. However, simultaneous broadband absorption and transmission are difficult to achieve in simple metallic nano-slit arrays due to small number of possible resonances that could be excited in the metallic elements.

Here, we propose ultra-broadband polarizers, reflective and absorptive, in the visible range using multi-scaled metallic (Ag) nano wire-slit arrays which could be realized by using a two-times shadowing vapor deposition method [27]. In our design, instead of 2D configuration, the multi-scaled metallic elements are arranged in 3D with specific arrangements so as to achieve high and broadband transmission. We show first in our simulations the transition from reflective to absorptive type as more metallic elements, e.g. Ag wires/strips, of different lengths/widths are incorporated into the nano arrays. For the absorptive type array, averaged (from 500 – 800nm) transmittance 3% and 71% for X and Y polarizations, respectively, can be achieved. More importantly, simultaneous high absorption, ~81%, and low reflection, 15%, for X-polarization incidence are feasible. Then, by using a two-times shadowing vapor deposition method, we fabricated the Ag nano wire-slit arrays exhibiting large broadband transmission and absorption differences for linearly polarized incident light, in good agreement with simulations. The large optical responses are due to the collective electron oscillations in the Ag wires/strips along X-polarization direction at different wavelengths. Moreover, reflectance of both X- and Y-polarizations are very low (almost zero for specific resonant wavelengths) across a broad spectrum and thus the Ag nano arrays can also be used as anti-reflective materials in the photovoltaic devices.

We started from a simple array of rectangular metallic wires along the horizontal X-direction and introduced more length scales into the array by reducing the lengths of selected wires and also by inserting metallic strips along the vertical Y-direction as shown in the schematics in the third column of Fig. 1 for four models, labelled I to IV. Model I, consisting of parallel long wires, has a similar configuration as the nano-slit array reported earlier [26]. Model II has alternate long (continuous) wires replaced by short wires. Model III has an extra vertical bottom Ag strip connecting the top long Ag wires in Model II. (Note that now the bottom vertical Ag strips are not on the same plane as the long Ag wires due to the shadowing vapor deposition used in the fabrication as explained in below.) Model IV has one more short horizontal Ag wire and also another bottom vertical Ag strip as in Model III.

We then carried out numerical simulations for the different models of the Ag nano wire-slit arrays as shown in Fig. 1 using commercial finite-integration time-domain algorithm (CST Microwave Studio) software to obtain optical responses of the arrays in the visible range. We used the dielectric constant of shadowing deposited Ag extracted following the procedures as reported earlier [28, 29] and also known values for the PMMA and ITO glass to calculate the optical responses (transmittance $T$, reflectance $R$, and absorption $A=1-T-R$) for the arrays with parameters listed in Fig. 1 for linearly polarized light propagating along Z direction from the substrate side.

Figures 1(a)-(h) show transmittance, reflectance and absorption of linearly polarized incident light, X-polarized first column and Y-polarized second column, for the corresponding models of the Ag nano wire-slit arrays in the third column. The simulated transmittance for Model I, blue lines in Figs. 1(a) and (e), exhibits flat and ultra-broadband responses with large difference between X- (~0.07) and Y- (~0.9) polarizations. Furthermore, the reflectance, green lines in Figs. 1(a) and (e), shows complementary responses for the two polarizations, i.e. ~0.8 for X- and ~0.04 for Y- polarization. However, the absorption, red lines in Figs. 1(a) and (e), is relatively small. The high reflection for polarization parallel to the wires is due to large impedance mismatch at the metal-air interface, similar to that of metallic films [30, 31]. However, absorption of light in metal is due to coherent electron oscillations with the incident light, i.e. excitation of LSPR. For X-polarization incidence, the excitation in the long wires is weak for the visible range and thus absorption is small (~0.1), red line in Fig. 1(a). Moreover, for Y-polarization incidence, due to the small width (50 nm) of the metal wire there is no excitation for the visible range and thus absorption is even smaller (~0.05), red dash line in Fig. 1(e). As a result, most of the light can propagate through the array over a broad bandwidth due to the weak slit-resonance [26], blue dash line in Fig. 1(e). Our result confirms that Model I of our Ag nano wire-slit array can be used as an ultra-broadband reflective type polarizer in visible range.

Figures 1(b) and (f) show the results for Model II Ag nano wire-slit array. It is noted that the responses are not that much different from that of Model I nano array except that slightly larger absorption (~0.2) and smaller reflection (~0.7) are observed at long wavelengths for X-polarized incident light. This is due to the plasmonic resonance of the short Ag wire length $d-w_1$ in Model II, in addition to the responses in Model I. As for the Y-polarization incidence, there is practically no difference between Model I and II nano arrays. Thus Model II nano array can also be used as broadband reflective polarizer.

Figures 1(c) and (g) show the responses for Model III Ag nano wire-slit array. It is clear that by adding a bottom vertical Ag strip, width $w_1$ and connecting the top horizontal long Ag wires, the responses are drastically different from that of Model II nano array. Now there is an obvious resonance associated with the bottom vertical Ag strip which is manifested as large absorption with peak value ~0.9, red line in Fig. 1(c), around 600 nm for X-polarized incident light. Correspondingly, the reflectance also shows a trough with minimum value ~0.08 at 600 nm. The absorption (~0.6) is mainly due to dissipations of horizontal current flows in the bottom Ag strips around 600 nm and is broadened to longer wavelengths by the extra current oscillations in the Ag short wires and also in the side walls of the long Ag wires connected to the bottom Ag strips. Despite the opposite behavior of the absorption and reflection around 600 nm, the transmittance (~0.05) is similar to that of Model II nano array. In comparison, for Y-polarization incidence, the transmittance, blue dash line in Fig. 1(g), is still large (~0.8) and flat, and the reflectance and absorption are still low because of the weak slit-resonance. Overall, the Model III Ag nano wire-slit array behaves as a broadband absorptive linear polarizer.

In order to further increase the absorption bandwidth of the Ag nano wire-slit array, another vertical Ag strip is added to Model III to form Model IV nano array as shown in Fig. 1. As a result, there are now three important length scales: $w_2$, $w_3$, and $w_2+w_3$, leading to three dominant resonances at 533, 598, and 680 nm as shown in Fig. 1(d). Note that, despite these resonances the absorption is still relatively flat and large ~0.8 averaged over all wavelengths for X-polarization incident light, Fig. 1(d), because of extra excitations of plasmonic resonances associated with the multi-scaled Ag side walls and short wires. Furthermore, the absorption is also flat and relatively small ~0.22 for Y-polarization incident light, Fig. 1(h). In contrast, the reflectance is now small, green lines in Figs. 1(d) and (h), ~0.15 for X- (much smaller than that of model III ~0.34, green curve in Fig. 1(c), and ~0.07 for Y-polarizations. However, there is still a large difference in the transmittance for X- (Fig. 1(d)) and Y- (Fig. 1(h)) polarizations. Combining all the responses, Model IV nano array exhibits ultra-broadband large transmittance ~0.71 for Y-polarization incidence but small ~0.03 for X-polarization. Thus Model IV nano array can serve as an ultra-broadband absorptive polarizer. In principle, one could keep adding more length scales into the array by replacing the long Ag wires in Model I by shorter ones to achieve smoother and flatter responses. However, this will reduce the total transmittance because of larger absorptions due to more plasmonic resonances. Nevertheless, the multi-scaled Ag nano wire-slit array could be optimized for specific bandwidth and absorption applications.

It is well known that plasmonic resonance on metal surfaces depends very much on the size of the metal. The absorption peak, e.g. shown in Fig. 1(c), is tied to the size (width $w_1$) of the bottom vertical Ag strip. Thus varying its width is a simple and effective way to control the resonance. Figure 2(a) shows the reflectance and absorption for three bottom Ag strip widths, $w_1$ = 60, 80, and 100 nm, of the Model III Ag nano wire-slit array for X-polarization incidence. It is clear that the resonance peak is red-shifted with increasing FWHM as the width increases. Furthermore, the absorption peak wavelength has almost a linear dependency on $w_1$ with roughly slow varying reflectance dip (0.002-0.12) and absorption peak (0.83-0.96) as shown in Fig. 2(b). To further characterize the Model III nano array, Figures 2(c)-(e) show the averaged (from 500 to 850 nm) transmittance, reflectance, and absorption for X- (solid symbols) and Y- (open symbols) polarization incidence, respectively. For X-polarization incidence, the averaged transmittance is very low and increases very slowly from 0.01 to 0.12 with $w_1$, solid blue squares in Fig. 2(c). In contrast, the averaged reflectance first decreases significantly with $w_1 <$ ~85 nm and then gently for $w_1 >$ 85 nm as shown by the solid green triangles in Fig. 2(d) whereas the absorption first increases with $w_1$, peaks (A ~0.6) at $w_1$ ~ 85 nm, and then decreases slowly for $w_1 >$ 85 nm, solid red circles in Fig. 2(e). It is now clear that the Model III nano array can be optimized for best absorption and bandwidth by tuning the bottom vertical Ag strip width to achieve roughly a maximum averaged absorption ~0.6 for $w_1$ ~85 nm. On the contrary, for Y-polarization incidence, open symbols in Figs. 2(c)-(e), the averaged transmittance (~0.81), reflectance (~0.06), and absorption (~0.13) do not change much with $w_1$. The above analysis can also be applied to Model IV Ag nano array with two bottom vertical Ag strips to achieve flatter and larger absorption and transmission for a specific polarization direction as an ultra-broadband absorptive polarizer.

As a show principle, two Ag nano wire-slit arrays, Model I and III, were fabricated by using a two-times shadowing vapor deposition method reported earlier [27] as shown in schematics in Figs. 3(a) and (b), respectively. The corresponding SEM images of the fabricated Ag arrays are shown in Figs. 3(c) and 3(d). The fabrication was straight forward. To begin templates of PMMA array with multi-length parallel strips/barriers on ITO glass were prepared by standard e-beam lithography (EBL) and lift-off procedures. Then, Ag was coated onto the PMMA strips by a shadowing vapor deposition method in two directions indicated by the orange arrows in the Y-Z plane and 45$^o$ with the Z-axis, see inset in Figs. 3(a) and (b). As a result, 30 nm and 15 nm thick Ag, shown as orange color in Figs. 3(a) and (b), was coated onto the top and the side walls of the PMMA strips, respectively, to form the Ag nano wire-slit array. Note that the side walls of the Model I nano array, Fig. 3(a), are not in contact with the substrate due to the shadowing effect. However as for the Model III nano array, a 15 nm thick Ag strip was now coated onto the substrate along the Y-(vertical) direction, shown clearly in the normal SEM image in the inset of Fig. 3(d), because the PMMA strips could not stop the Ag flux from reaching the substrate surface. (Note that the bottom Ag strip has a small doubly-coated region in the middle, inset in Fig. 3(d), due to the two depositions in opposite directions.) As a result, the long horizontal (X-direction) Ag wires are now connected to each other by the bottom vertical (Y-direction) Ag strips to a 3D network. Note that the short horizontal Ag wires are still disconnected from this network.

We characterized the optical responses of the Model I and III Ag nano wire-slit arrays by measuring the transmittance and reflectance, and hence the absorption, using a visible range (covering 500 to 850 nm wavelengths) microscopic optical setup reported before with linearly polarized normal incidence from the substrate side propagating along Z direction [27, 29]. Figures 4(d)-(f) and 4(j)-(l) show the experimental transmittance, reflectance, and absorption for the Model I and Model III nano arrays, respectively. The transmittance is roughly flat, small ~0.12 for X-polarization but large ~ 0.7 for Y-polarization incident light, for both nano arrays. However, while the reflectance (Fig. 4(e)) of the Model I nano array is relatively flat (~0.7 for X-polarization and ~0.13 for Y-polarization), the reflectance (Fig. 4(k)) of the Model III Ag array exhibits a broad trough around 600 nm for the X-polarization incident light. The reflection trough corresponds well to the large absorption peak for the X-polarization incidence as shown clearly in Fig. 4(l). Overall, the experimental results agree very well with the simulation results for the two Ag nano arrays, Figs. 4(a)-(c) and Figs. 4(g)-(i). Our experiment verifies that Model I Ag nano array, having a flat and very different transmittance and reflectance for different incident polarizations, can be used as an ultra-broadband reflective polarizer as expected for simple parallel conducting wires [25, 26] while Model III Ag nano array, having strong and broad absorption, can be used as a broadband absorptive polarizer.

To conclude, plasmonic multi-scaled metallic nano wire-slit arrays have been designed as ultra-broadband reflective and absorptive polarizers with large transmittance difference for orthogonally linearly polarized incident light in the visible range. The bandwidth of absorption can be tuned by incorporating more wires/strips with different lengths/widths to the arrays. We have also realized, by using a simple two-times shadowing vapor deposition, two types of Ag nano wire-slit array exhibiting optical responses in good agreement with simulations. Our multi-scaled nano wire-slit array design can find applications in broadband linear polarizers and anti-reflective materials in both optics and photovoltaics.

ACKNOWLEDGMENT

Support from Hong Kong RGC grants HKUST2 CRF 11G and AoE P-02/12 is gratefully acknowledged. The technical support of the Raith-HKUST Nanotechnology Laboratory for the electron-beam lithography facility at MCPF (SEG_HKUST08) is hereby acknowledged. The simulations were carried out using the server in the Key Laboratory of Advanced Micro-structure Materials, Ministry of Education, China; and also the School of Physics Science and Engineering, Tongji University in collaboration with Prof. Y. Li in Tongji University.

References:

[1] W. L. Barnes, "Metallic metamaterials and plasmonics", *Phil. Trans. R. Soc. A,* 369 (2011).

[2] W. L. Barnes, A. Dereux, and T. W. Ebbesen, "Surface plasmon subwavelength optics", *nature*, **424**, 824 (2003).

[3] E. Hutter and J. H. Fendler, "Exploitation of localized surface plasmon resonance"*, Adv. Mater*. **16**, 1685 (2004).

[4] Y. Cui, J. Xu, K. H. Fung, Y. Jin, A. Kumar, S. He, and N. X. Fang, "A thin film broadband absorber based on multi-sized nanoantennas"*, Appl. Phys. Lett.* **99**, 253101 (2011).

[5] J. A. Schuller, E. S. Barnard, W. Cai, Y. C. Jun, J. S. White, and M. L. Brongersma, "Plasmonics for extreme light concentration and manipulation", *Nat. Mater.* **9**, 193 (2010).

[6] M. E. Stewart, C. R. Anderton, L. B. Thompson, J. Maria, S. K. Gray, J. A. Rogers, and R. G. Nuzzo, "Nanostructured Plasmonic Sensors", *Chem. Rev.*, **108**, 494 (2008).

[7] J. Dorfmuller, R. Vogelgesang, W. Khunsin, C. Rockstuhl, C. Etrich, and K. Kern, "Plasmonic Nanowire Antennas: Experiment, Simulation, and Theory", *Nano Lett.* **10**, 3596 (2010).

[8] A. Bouhelier, J. Renger, M. R. Beversluis, and L. Novotny, "Plasmon-coupled tip-enhanced near-field optical microscopy", *J. Microsc.* **210**, 220 (2003).

[9] N. Liu, M. Mesch, T. Weiss, M. Hentschel, and H. Giessen, "Infrared Perfect Absorber and Its Application As Plasmonic Sensor", *Nano Lett.* **10**, 2342 (2010).

[10] C. W. Cheng, M. N. Abbas, C. W. Chiu, K. T. Lai, M. H. Shih, and Y. C. Chang, "Wide-angle polarization independent infrared broadband absorbers based on metallic multisized disk arrays", *Opt. Express.* **20**, 10376 (2012).

[11] B. Zhang, J. Hendrickson, and J. Guo, "Multispectral near-perfect metamaterial absorbers using spatially multiplexed plasmon resonance metal square structures", *J. Opt. Soc. Am. B* **30**, 656 (2013).

[12] R. Feng, W. Ding, L. Liu, L. Chen, J. Qiu, and G. Chen, "Dual-band infrared perfect absorber based on asymmetric T-shaped plasmonic array", *Opt. Express*, **22**, 335 (2014).

[13] N. I. Landy, S. Sajuyigbe, J. J. Mock, D. R. Smith, and W. J. Padilla, "Perfect Metamaterial Absorber", *Phys. Rev. Lett.* **100**, 207402 (2008).

[14] Y. Cui, K. H. Fung, J. Xu, H. Ma, Y. Jin, S. He, and N. X. Fang, "Ultrabroadband Light Absorption by a Sawtooth Anisotropic Metamaterial Slab", *Nano Lett.* **12**, 1443 (2012).

[15] H. Wang, K. O'Dea, and L. Wang, "Selective absorption of visible light in film-coupled nanoparticles by exciting magnetic resonance", *Opt. Lett.* **39**, 1457 (2014).

[16] Y. Zhao and A. Alu, "Tailoring the Dispersion of Plasmonic Nanorods To Realize Broadband Optical Meta-Waveplates", *Nano Lett.* **13**, 1086 (2013)

[17] F. I. Baida, M. Boutria, R. Oussaid, and D. Van Labeke, "Enhanced-transmission metamaterials as anisotropic plates", *Phys. Rev. B.* **84**, 035107 (2011).

[18] E. H. Khoo, E. P. Li, and K. B. Crozier, "Plasmonic wave plate based on subwavelength nanoslits", *Opt. Lett.* **36**, 2498 (2011).

[19] M. A. Katsa, P. Geneveta, G. Aousta, N. Yua, R. Blancharda, F. Aietaa, Z. Gaburroa, and F. Capasso, "Giant birefringence in optical antenna arrays with widely tailorable optical anisotropy", *Proc. Natl. Acad. Sci. USA*, **109**, 12364 (2012).

[20] W. Gotschy, K. Vonmetz, A. Leitner, and F. R. Aussenegg, "Optical dichroism of lithographically designed silver nanoparticle films", *Opt. Lett.* **21,** 1099 (1996).

[21] R. Gordon, A. G. Brolo, A. McKinnon, A. Rajora, B. Leathem, and K. L. Kavanagh, "Strong Polarization in the Optical Transmission through Elliptical Nanohole Arrays", *Phys. Rev. Lett.* **92**, 037401 (2004).

[22] J. Elliott, I. I. Smolyaninov, N. I. Zheludev, and A. V. Zayats, "Wavelength dependent birefringence of surface plasmon polaritonic crystals", *Phys. Rev. B.* **70**, 233403 (2004).

[23] P. Biagioni, J. S. Huang, L. Duo, M. Finazzi, and B. Hecht1, "Cross Resonant Optical Antenna", *Phys. Rev. Lett.* **102**, 256801 (2009).

[24] N. W. Schuler and L. Mass, "Iodine stained light polarizer", US4166871 A, 1979.

[25] E. Hecht and C. Zhang, Optics, Fourth Edition, High education press, 2005.

[26] J. Zhou and L. J. Guo, "Transition from a spectrum filter to a polarizer in a metallic nano-slit array", *Sci. Rep.* **4**, 3614 (2014).

[27] C. Han, H. M. Leung, and W. Y. Tam, "Chiral metamaterials by shadowing vapor deposition", *J. Opt.* **15,** 072101 (2013).

[28] H. M. Leung, C. Han, Y. Li, C. T. Chan, and W. Y. Tam, "Modeling quasi-3D chiral metamaterials fabricated by shadowing vapor deposition", *J. Opt.* **16**, 015102 (2013).

[29] C. Han and W. Y. Tam, "Chirality from shadowing deposited metallic nanostructure". *Phot. Nano. Fund. Appl.* doi:10.1016/j.photonics.2014.10.002, (2014).

[30] D. A Weston, "Electromagnetic Compatibility: Principles and Applications", Second Edition, CRC press, p283 (2001).

[31] D. J. Brady, "Optical Imaging and Spectroscopy", John Wiley & Sons, p154 (1961).

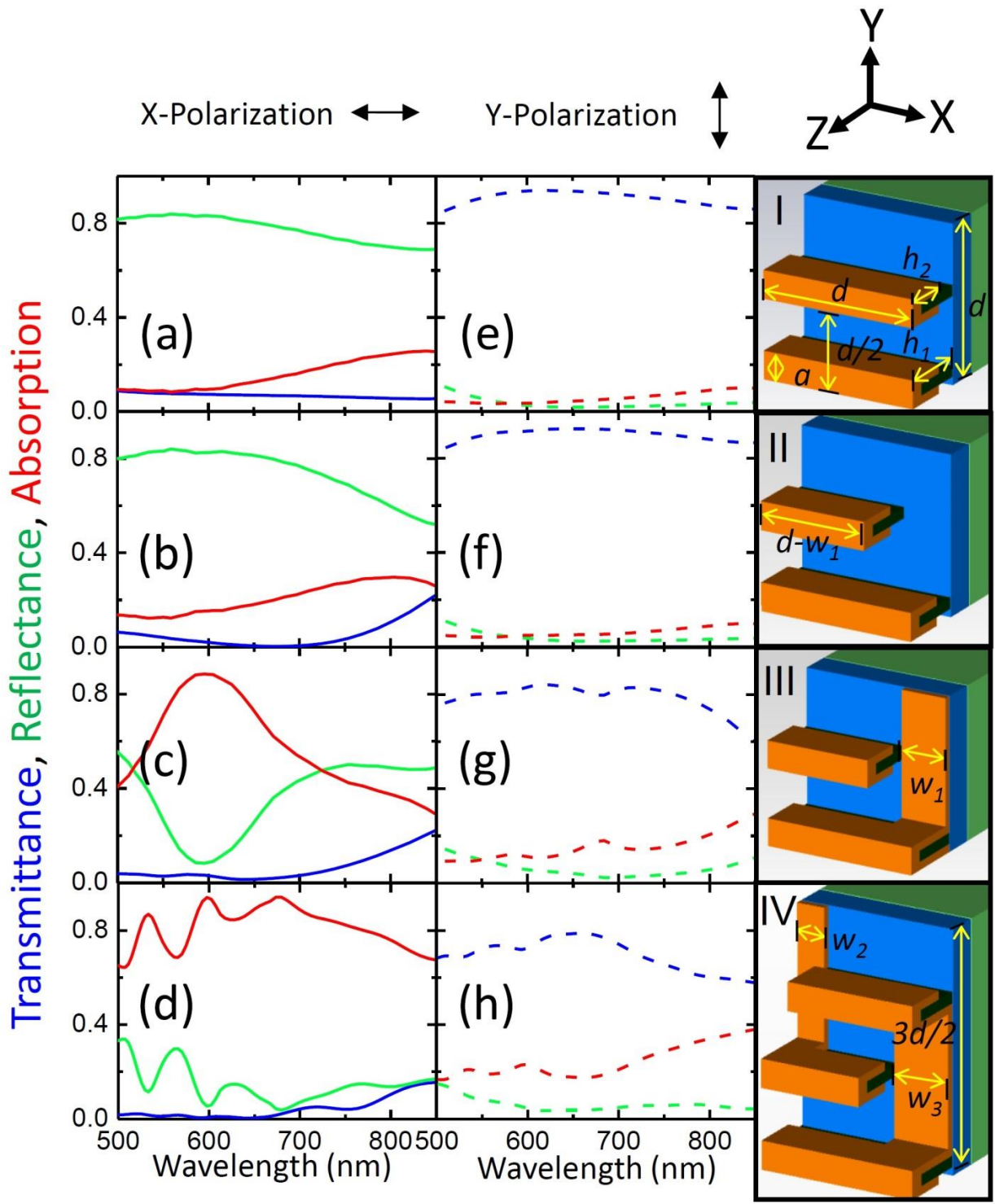

Fig. 1: (a)-(h) Calculated transmittance (blue lines), reflectance (green lines) and absorption (red lines) of Ag nano wire-slit arrays, models I-IV shown as schematics in the third column, for X- (solid lines in first column) and Y-polarization (dashed lines in second column) incident light. Color codes for the schematics are: Ag orange, PMMA green, ITO blue, and substrate glass light-green. The parameters for the models are: $d$=280 nm, $a$=50 nm, $h_1$=130 nm, $h_2$=85 nm, $w_1$= 85 nm, $w_2$=50 nm, $w_3$=100 nm.

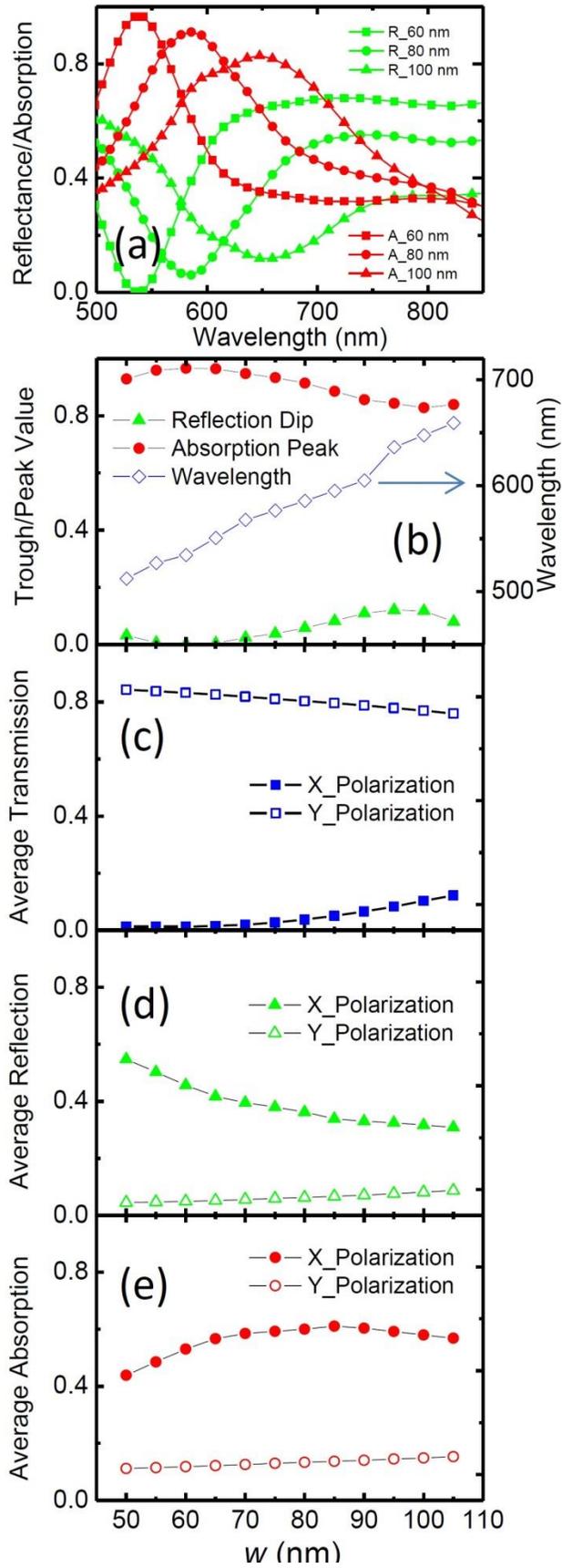

Fig. 2: (a)-(e) Simulation results for Model III Ag nano wire-slit array in Fig. 1. (a) Calculated reflectance (green lines) and absorption (red lines) for different bottom Ag strip widths for X-polarization incidence. (b) Reflection dip (green solid triangles) and absorption peak (red solid circles), and corresponding wavelength (blue open squares), as a function of the bottom Ag strip width $w$ for X-polarization incidence. (c)-(e) Calculated (averaged over 500-850 nm) transmittance, reflectance, and absorption as a function of the bottom Ag strip width $w$ for X- (solid symbols) and Y- (open symbols) polarization incidence, respectively.

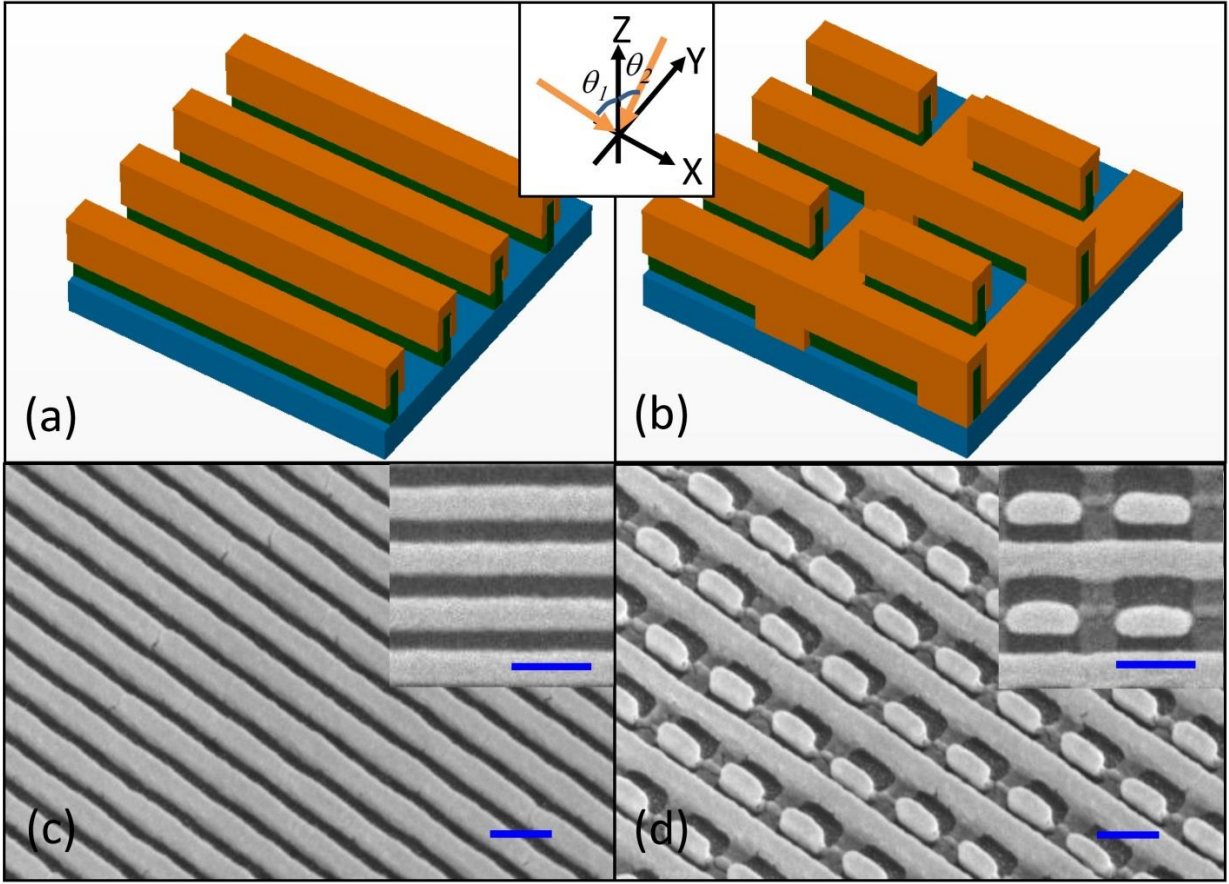

Fig. 3: (a) and (b) Schematics of Model I and III Ag nano wire-slit arrays. The dimensions for (a) and (b) are the same as in Fig. 1. Orange arrows indicate the directions of Ag flux with $\theta_1 = \theta_2 = 45°$ for the two-times shadowing vapor deposition. (c) and (d) Tilted SEM images for Model I and III Ag nano wire-slit arrays, respectively. Insets are expanded normal SEM images. Blue scale bars are 200 nm.

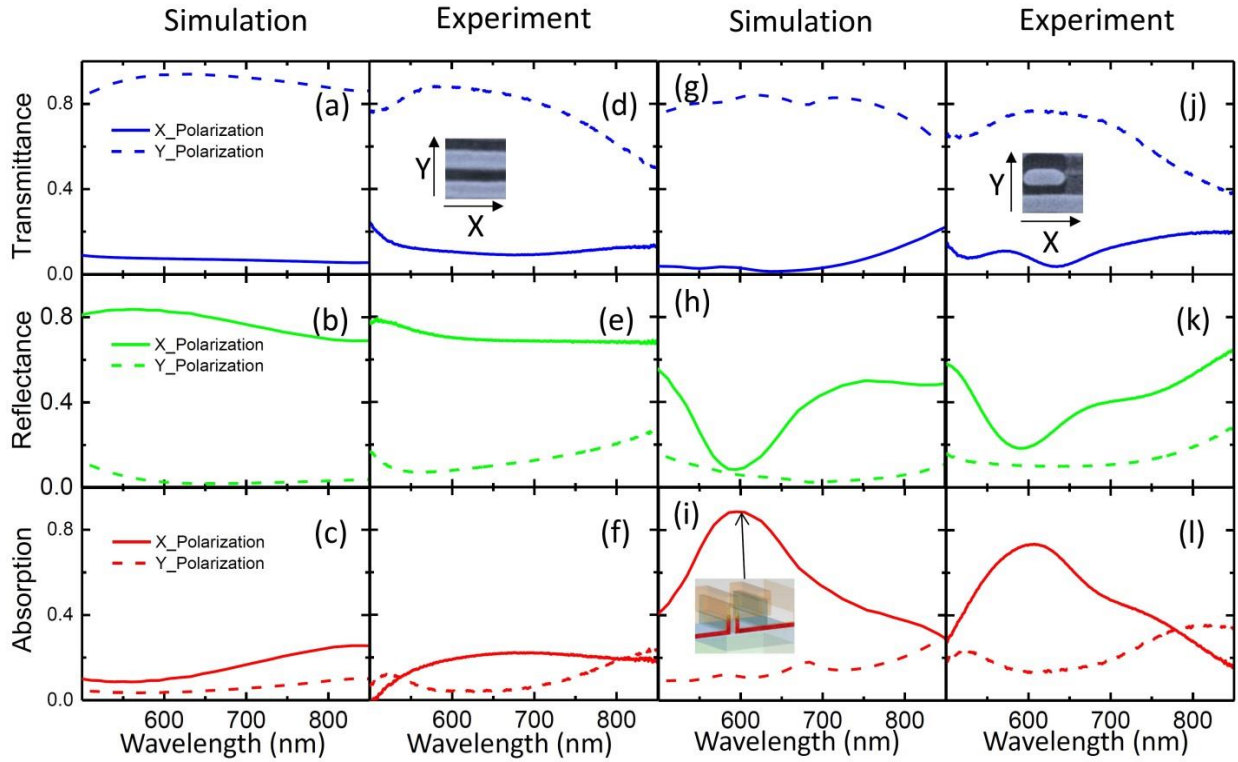

Fig. 4: Transmittance (blue lines), reflectance (green lines) and absorption (red lines) of the Model I (first and second columns) and Model III (third and fourth columns) Ag nano wire-slit arrays. Solid lines and dashed lines are X- and Y-polarization incidence, respectively. The first and third columns are the simulation results; the second and fourth columns are the experimental results. Inset in (i) is current distribution for X-polarized incident light at 600 nm. Red color represents high current density.